\documentclass[conference]{IEEEtran}
\usepackage[utf8]{inputenc}      
\usepackage[T1]{fontenc}         
\usepackage[english]{babel}      
\usepackage{graphicx}            
\usepackage{cite}                
\usepackage[scaled]{helvet} 
\usepackage[bookmarks=false]{hyperref}            
\usepackage{comment}             

\title{Explainable Queries over Event Logs}

\author{%
\IEEEauthorblockN{Sylvain Hallé%
}
\\
\IEEEauthorblockA{%
Laboratoire d'informatique formelle\\
Université du Québec à Chicoutimi, Canada\\
}

}

\usepackage{amsmath,amsfonts,amssymb}

\usepackage{subfig}

\usepackage{paralist}

\usepackage{listings}
\usepackage{xcolor}
\usepackage{todonotes}

\newcommand{\palette}[1]{\textsf{#1}}
\newcommand{\ev}[1]{\textsl{#1}}

\usepackage{newtxtext,newtxmath} 
\usepackage{microtype}       
\graphicspath{{fig/}}

\begin{document}

\maketitle
\begin{abstract}
Added value can be extracted from event logs generated by business processes in various ways. 
However, although complex computations can be performed over event logs, the result of such computations is often difficult to explain; in particular, it is hard to determine what parts of an input log actually matters in the production of that result. This paper describes how an existing log processing library, called BeepBeep, can be extended in order to provide a form of provenance: individual output events produced by a query can be precisely traced back to the data elements of the log that contribute to (i.e.\ ``explain'') the result.

\end{abstract}

\makeatletter
\let\l@ENGLISH\l@english
\makeatother



\newcommand{\maxOhMemory}{\href{M1.0}{4300.0}}

\newcommand{\maxOhThroughput}{\href{M1.1}{21.7}}

\newcommand{\machinestring}{\href{M3.0}{Intel CORE i5-7200U 2.5 GHz running Ubuntu 18.04}}

\newcommand{\jvmram}{\href{M3.1}{1746}}

\newcommand{\numexperiments}{\href{M3.2}{6}}

\newcommand{\numdatapoints}{\href{M3.3}{36}}

\newcommand{\maxLines}{\href{M2.0}{86 million}}

%

\newsavebox{\ttCompTime}
\begin{lrbox}{\ttCompTime}
\begin{tabular}{|c|c|c|}
\hline
\textbf{Query} & \textbf{No tracker (Hz)} & \textbf{With tracker (Hz)}\\
\hline\hline
{\href{T1.2.0}{LTL property}} & {\href{T1.2.1}{9452.741}} & {\href{T1.2.2}{2128.3252}}\\
\hline
{\href{T1.1.0}{Process lifecycle}} & {\href{T1.1.1}{4283.0835}} & {\href{T1.1.2}{2099.727}}\\
\hline
{\href{T1.0.0}{Window product}} & {\href{T1.0.1}{333366.66}} & {\href{T1.0.2}{15386.154}}\\

\hline
\end{tabular}
\end{lrbox}

\newsavebox{\tCompTimeVs}
\begin{lrbox}{\tCompTimeVs}
\begin{tabular}{|c|c|}
\hline
\textbf{No tracker (Hz)} & \textbf{With tracker (Hz)}\\
\hline\hline
{\href{T2.1.0}{4283.0835}} & {\href{T2.1.1}{2099.727}}\\
\hline
{\href{T2.2.0}{9452.741}} & {\href{T2.2.1}{2128.3252}}\\
\hline
{\href{T2.0.0}{333366.66}} & {\href{T2.0.1}{15386.154}}\\

\hline
\end{tabular}
\end{lrbox}

\newsavebox{\ttCompMem}
\begin{lrbox}{\ttCompMem}
\begin{tabular}{|c|c|c|}
\hline
\textbf{Query} & \textbf{No tracker (B)} & \textbf{With tracker (B)}\\
\hline\hline
{\href{T3.2.0}{LTL property}} & {\href{T3.2.1}{12341}} & {\href{T3.2.2}{53027241}}\\
\hline
{\href{T3.1.0}{Process lifecycle}} & {\href{T3.1.1}{24039551}} & {\href{T3.1.2}{40353151}}\\
\hline
{\href{T3.0.0}{Window product}} & {\href{T3.0.1}{5294}} & {\href{T3.0.2}{7404930}}\\

\hline
\end{tabular}
\end{lrbox}

\newsavebox{\tCompMemVs}
\begin{lrbox}{\tCompMemVs}
\begin{tabular}{|c|c|}
\hline
\textbf{No tracker (B)} & \textbf{With tracker (B)}\\
\hline\hline
{\href{T4.0.0}{5294}} & {\href{T4.0.1}{7404930}}\\
\hline
{\href{T4.2.0}{12341}} & {\href{T4.2.1}{53027241}}\\
\hline
{\href{T4.1.0}{24039551}} & {\href{T4.1.1}{40353151}}\\

\hline
\end{tabular}
\end{lrbox}

\newsavebox{\tMemPerEvent}
\begin{lrbox}{\tMemPerEvent}
\begin{tabular}{|c|c|}
\hline
\textbf{Query} & \textbf{Memory per event}\\
\hline\hline
{\href{T5.0.0}{LTL property}} & {\href{T5.0.1}{5300}}\\
\hline
{\href{T5.2.0}{Process lifecycle}} & {\href{T5.2.1}{1631}}\\
\hline
{\href{T5.1.0}{Window product}} & {\href{T5.1.1}{739}}\\

\hline
\end{tabular}
\end{lrbox}

\newsavebox{\ttMemQueryWindowPproduct}
\begin{lrbox}{\ttMemQueryWindowPproduct}
\begin{tabular}{|c|c|c|}
\hline
\textbf{Length} & \textbf{With tracker} & \textbf{No tracker}\\
\hline\hline
{\href{T7.0.0}{0}} & {\href{T7.0.1}{0}} & {\href{T7.0.2}{0}}\\
\hline
{\href{T7.1.0}{1000}} & {\href{T7.1.1}{744930}} & {\href{T7.1.2}{5294}}\\
\hline
{\href{T7.2.0}{2000}} & {\href{T7.2.1}{1484930}} & {\href{T7.2.2}{5294}}\\
\hline
{\href{T7.3.0}{3000}} & {\href{T7.3.1}{2224930}} & {\href{T7.3.2}{5294}}\\
\hline
{\href{T7.4.0}{4000}} & {\href{T7.4.1}{2964930}} & {\href{T7.4.2}{5294}}\\
\hline
{\href{T7.5.0}{5000}} & {\href{T7.5.1}{3704930}} & {\href{T7.5.2}{5294}}\\
\hline
{\href{T7.6.0}{6000}} & {\href{T7.6.1}{4444930}} & {\href{T7.6.2}{5294}}\\
\hline
{\href{T7.7.0}{7000}} & {\href{T7.7.1}{5184930}} & {\href{T7.7.2}{5294}}\\
\hline
{\href{T7.8.0}{8000}} & {\href{T7.8.1}{5924930}} & {\href{T7.8.2}{5294}}\\
\hline
{\href{T7.9.0}{9000}} & {\href{T7.9.1}{6664930}} & {\href{T7.9.2}{5294}}\\
\hline
{\href{T7.10.0}{10000}} & {\href{T7.10.1}{7404930}} & {\href{T7.10.2}{5294}}\\

\hline
\end{tabular}
\end{lrbox}

\newsavebox{\ttTpQueryWindowPproduct}
\begin{lrbox}{\ttTpQueryWindowPproduct}
\begin{tabular}{|c|c|c|}
\hline
\textbf{Length} & \textbf{With tracker} & \textbf{No tracker}\\
\hline\hline
{\href{T9.0.0}{0}} & {\href{T9.0.1}{0}} & {\href{T9.0.2}{0}}\\
\hline
{\href{T9.1.0}{1000}} & {\href{T9.1.1}{30}} & {\href{T9.1.2}{4}}\\
\hline
{\href{T9.2.0}{2000}} & {\href{T9.2.1}{58}} & {\href{T9.2.2}{7}}\\
\hline
{\href{T9.3.0}{3000}} & {\href{T9.3.1}{129}} & {\href{T9.3.2}{9}}\\
\hline
{\href{T9.4.0}{4000}} & {\href{T9.4.1}{179}} & {\href{T9.4.2}{13}}\\
\hline
{\href{T9.5.0}{5000}} & {\href{T9.5.1}{229}} & {\href{T9.5.2}{16}}\\
\hline
{\href{T9.6.0}{6000}} & {\href{T9.6.1}{299}} & {\href{T9.6.2}{19}}\\
\hline
{\href{T9.7.0}{7000}} & {\href{T9.7.1}{372}} & {\href{T9.7.2}{22}}\\
\hline
{\href{T9.8.0}{8000}} & {\href{T9.8.1}{473}} & {\href{T9.8.2}{25}}\\
\hline
{\href{T9.9.0}{9000}} & {\href{T9.9.1}{553}} & {\href{T9.9.2}{27}}\\
\hline
{\href{T9.10.0}{10000}} & {\href{T9.10.1}{650}} & {\href{T9.10.2}{30}}\\

\hline
\end{tabular}
\end{lrbox}

\newsavebox{\ttMemQueryProcessPlifecycle}
\begin{lrbox}{\ttMemQueryProcessPlifecycle}
\begin{tabular}{|c|c|c|}
\hline
\textbf{Length} & \textbf{With tracker} & \textbf{No tracker}\\
\hline\hline
{\href{T11.0.0}{0}} & {\href{T11.0.1}{0}} & {\href{T11.0.2}{0}}\\
\hline
{\href{T11.1.0}{1000}} & {\href{T11.1.1}{4010709}} & {\href{T11.1.2}{2381285}}\\
\hline
{\href{T11.2.0}{2000}} & {\href{T11.2.1}{8170555}} & {\href{T11.2.2}{4894283}}\\
\hline
{\href{T11.3.0}{3000}} & {\href{T11.3.1}{12173129}} & {\href{T11.3.2}{7271705}}\\
\hline
{\href{T11.4.0}{4000}} & {\href{T11.4.1}{16185217}} & {\href{T11.4.2}{9653969}}\\
\hline
{\href{T11.5.0}{5000}} & {\href{T11.5.1}{20339607}} & {\href{T11.5.2}{12166967}}\\
\hline
{\href{T11.6.0}{6000}} & {\href{T11.6.1}{24444617}} & {\href{T11.6.2}{14631545}}\\
\hline
{\href{T11.7.0}{7000}} & {\href{T11.7.1}{28452295}} & {\href{T11.7.2}{17008967}}\\
\hline
{\href{T11.8.0}{8000}} & {\href{T11.8.1}{32325787}} & {\href{T11.8.2}{19265339}}\\
\hline
{\href{T11.9.0}{9000}} & {\href{T11.9.1}{36359573}} & {\href{T11.9.2}{21671813}}\\
\hline
{\href{T11.10.0}{10000}} & {\href{T11.10.1}{40353151}} & {\href{T11.10.2}{24039551}}\\

\hline
\end{tabular}
\end{lrbox}

\newsavebox{\ttTpQueryProcessPlifecycle}
\begin{lrbox}{\ttTpQueryProcessPlifecycle}
\begin{tabular}{|c|c|c|}
\hline
\textbf{Length} & \textbf{With tracker} & \textbf{No tracker}\\
\hline\hline
{\href{T13.0.0}{0}} & {\href{T13.0.1}{0}} & {\href{T13.0.2}{0}}\\
\hline
{\href{T13.1.0}{1000}} & {\href{T13.1.1}{232}} & {\href{T13.1.2}{54}}\\
\hline
{\href{T13.2.0}{2000}} & {\href{T13.2.1}{429}} & {\href{T13.2.2}{147}}\\
\hline
{\href{T13.3.0}{3000}} & {\href{T13.3.1}{707}} & {\href{T13.3.2}{268}}\\
\hline
{\href{T13.4.0}{4000}} & {\href{T13.4.1}{963}} & {\href{T13.4.2}{432}}\\
\hline
{\href{T13.5.0}{5000}} & {\href{T13.5.1}{1261}} & {\href{T13.5.2}{671}}\\
\hline
{\href{T13.6.0}{6000}} & {\href{T13.6.1}{1594}} & {\href{T13.6.2}{929}}\\
\hline
{\href{T13.7.0}{7000}} & {\href{T13.7.1}{2078}} & {\href{T13.7.2}{1209}}\\
\hline
{\href{T13.8.0}{8000}} & {\href{T13.8.1}{2479}} & {\href{T13.8.2}{1530}}\\
\hline
{\href{T13.9.0}{9000}} & {\href{T13.9.1}{3947}} & {\href{T13.9.2}{1953}}\\
\hline
{\href{T13.10.0}{10000}} & {\href{T13.10.1}{4762}} & {\href{T13.10.2}{2334}}\\

\hline
\end{tabular}
\end{lrbox}

\newsavebox{\ttMemQueryLTLPproperty}
\begin{lrbox}{\ttMemQueryLTLPproperty}
\begin{tabular}{|c|c|c|}
\hline
\textbf{Length} & \textbf{With tracker} & \textbf{No tracker}\\
\hline\hline
{\href{T15.0.0}{0}} & {\href{T15.0.1}{0}} & {\href{T15.0.2}{0}}\\
\hline
{\href{T15.1.0}{1000}} & {\href{T15.1.1}{5319723}} & {\href{T15.1.2}{12311}}\\
\hline
{\href{T15.2.0}{2000}} & {\href{T15.2.1}{10627595}} & {\href{T15.2.2}{12311}}\\
\hline
{\href{T15.3.0}{3000}} & {\href{T15.3.1}{15924907}} & {\href{T15.3.2}{12311}}\\
\hline
{\href{T15.4.0}{4000}} & {\href{T15.4.1}{21220841}} & {\href{T15.4.2}{12341}}\\
\hline
{\href{T15.5.0}{5000}} & {\href{T15.5.1}{26529769}} & {\href{T15.5.2}{12341}}\\
\hline
{\href{T15.6.0}{6000}} & {\href{T15.6.1}{31839753}} & {\href{T15.6.2}{12341}}\\
\hline
{\href{T15.7.0}{7000}} & {\href{T15.7.1}{37137945}} & {\href{T15.7.2}{12341}}\\
\hline
{\href{T15.8.0}{8000}} & {\href{T15.8.1}{42428393}} & {\href{T15.8.2}{12341}}\\
\hline
{\href{T15.9.0}{9000}} & {\href{T15.9.1}{47727787}} & {\href{T15.9.2}{12311}}\\
\hline
{\href{T15.10.0}{10000}} & {\href{T15.10.1}{53027241}} & {\href{T15.10.2}{12341}}\\

\hline
\end{tabular}
\end{lrbox}

\newsavebox{\ttTpQueryLTLPproperty}
\begin{lrbox}{\ttTpQueryLTLPproperty}
\begin{tabular}{|c|c|c|}
\hline
\textbf{Length} & \textbf{With tracker} & \textbf{No tracker}\\
\hline\hline
{\href{T17.0.0}{0}} & {\href{T17.0.1}{0}} & {\href{T17.0.2}{0}}\\
\hline
{\href{T17.1.0}{1000}} & {\href{T17.1.1}{456}} & {\href{T17.1.2}{177}}\\
\hline
{\href{T17.2.0}{2000}} & {\href{T17.2.1}{794}} & {\href{T17.2.2}{212}}\\
\hline
{\href{T17.3.0}{3000}} & {\href{T17.3.1}{1096}} & {\href{T17.3.2}{262}}\\
\hline
{\href{T17.4.0}{4000}} & {\href{T17.4.1}{1442}} & {\href{T17.4.2}{343}}\\
\hline
{\href{T17.5.0}{5000}} & {\href{T17.5.1}{1751}} & {\href{T17.5.2}{404}}\\
\hline
{\href{T17.6.0}{6000}} & {\href{T17.6.1}{2129}} & {\href{T17.6.2}{582}}\\
\hline
{\href{T17.7.0}{7000}} & {\href{T17.7.1}{2603}} & {\href{T17.7.2}{702}}\\
\hline
{\href{T17.8.0}{8000}} & {\href{T17.8.1}{3194}} & {\href{T17.8.2}{805}}\\
\hline
{\href{T17.9.0}{9000}} & {\href{T17.9.1}{4179}} & {\href{T17.9.2}{935}}\\
\hline
{\href{T17.10.0}{10000}} & {\href{T17.10.1}{4698}} & {\href{T17.10.2}{1057}}\\

\hline
\end{tabular}
\end{lrbox}

%

\newsavebox{\pCompTimeVs}
\begin{lrbox}{\pCompTimeVs}
\href{P1.0}{\includegraphics[page=1,width=\linewidth]{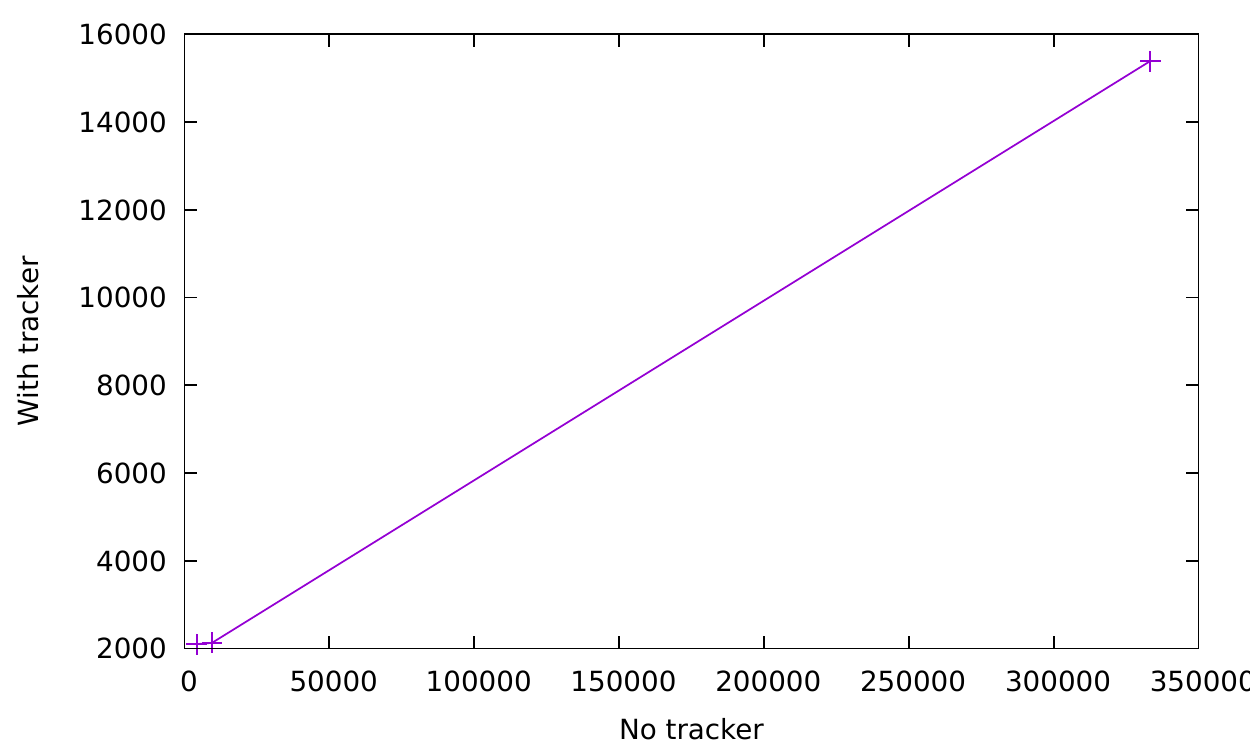}}
\end{lrbox}

\newsavebox{\pCompMemVs}
\begin{lrbox}{\pCompMemVs}
\href{P2.0}{\includegraphics[page=2,width=\linewidth]{labpal-plots.pdf}}
\end{lrbox}

\newsavebox{\pMemQueryWindowPproduct}
\begin{lrbox}{\pMemQueryWindowPproduct}
\href{P3.0}{\includegraphics[page=3,width=\linewidth]{labpal-plots.pdf}}
\end{lrbox}

\newsavebox{\pTpQueryWindowPproduct}
\begin{lrbox}{\pTpQueryWindowPproduct}
\href{P4.0}{\includegraphics[page=4,width=\linewidth]{labpal-plots.pdf}}
\end{lrbox}

\newsavebox{\pMemQueryProcessPlifecycle}
\begin{lrbox}{\pMemQueryProcessPlifecycle}
\href{P5.0}{\includegraphics[page=5,width=\linewidth]{labpal-plots.pdf}}
\end{lrbox}

\newsavebox{\pTpQueryProcessPlifecycle}
\begin{lrbox}{\pTpQueryProcessPlifecycle}
\href{P6.0}{\includegraphics[page=6,width=\linewidth]{labpal-plots.pdf}}
\end{lrbox}

\newsavebox{\pMemQueryLTLPproperty}
\begin{lrbox}{\pMemQueryLTLPproperty}
\href{P7.0}{\includegraphics[page=7,width=\linewidth]{labpal-plots.pdf}}
\end{lrbox}

\newsavebox{\pTpQueryLTLPproperty}
\begin{lrbox}{\pTpQueryLTLPproperty}
\href{P8.0}{\includegraphics[page=8,width=\linewidth]{labpal-plots.pdf}}
\end{lrbox}

\section{Introduction} 

Various kinds of information systems generate data streams in the form of sequences of data elements called \emph{event logs}. Sources of event logs are diverse: business process management engines, web servers, sensor networks, instrumented pieces of generic software can all be instructed to record information about their execution to a persistent storage medium. Added value can be extracted from event logs generated by these systems in various ways. Logs can be checked for compliance violations of best practices, adherence to predetermined sequences of events, detect deviations of some data point from a specified value, or be used to calculate various quality metrics. This process can take place after the system has completed its execution (\emph{offline} processing), or compute its results on-the-fly as the events from the source are ingested (\emph{streaming} processing). These two modes of operation are often grouped under the generic term ``event stream processing''.

Over the past decade, event stream processing systems have seen widespread use, with the advent of solutions such as Amazon Kinesis\footnote{\url{https://aws.amazon.com/kinesis}}, Apache Storm\footnote{\url{https://storm.apache.org}}, Flink\footnote{\url{https://flink.apache.org}}, Siddhi \cite{DBLP:conf/sc/SuhothayanGNCPN11} and Esper\footnote{\url{https://espertech.com}}. These systems provide rich processing capabilities, making it possible to evaluate complex queries over event logs. However, although intricate computations can be performed over these sources of data, the result of such computations is often difficult to explain. For example, a Flink pipeline can calculate some quality metric over instances of a process, and check that it always lies over some given threshold; however, if the result is false, how can one identify the source of the error? 

Developers of information systems in all disciplines are facing increasing pressure to come up with mechanisms to describe how a specific result is obtained --a concept called \emph{explainability}. Although the term is often tied to AI \cite{samek_explainable_2017}, explainability is desirable in other fields of computation. Hence, if a system fails to verify a given property, a counter-example is generally sought after as a means of understanding the source of the problem. This pressure often comes from regulations imposing constraints on the traceability of data processing, such as GDPR and BCBS. Yet, for most of the aforementioned engines, it is hard to determine what parts of an input log actually matters in the production of a given result. A user is typically left with the manual task of querying the log in various ways in order to investigate the reason for a surprising or irregular output result.

In Section \ref{sec:related}, we shall see that various technologies and frameworks have been developed over the years in order to provide a form of ``lineage'' or ``provenance'' information about the output of some computer system. However, none of these systems consider the special problem of explainability for event stream processing; in contrast, existing event stream processing systems provide very few in the way of lineage and explainability, leaving a gap that needs to be filled. In this paper, we describe how an existing log processing library, called BeepBeep \cite{beepbeep-book}, can be extended in order to provide a form of explanation mechanism: the output produced by a query can be precisely traced back to the individual data elements of the log that contribute to (i.e.\ ``explain'') the result.

Section \ref{sec:bb} shall introduce the basic concepts behind event stream processing in BeepBeep, and provide a few examples of simple queries that can be run on event logs. Section \ref{sec:lineage} describes the data lineage mechanism that has been added to the library as part of this work. This mechanism leverages the fact that calculations in BeepBeep are done by composing basic computation units together into event pipelines; therefore, in order to obtain end-to-end provenance, it suffices to define simple input/output relationships for each of these units separately. Finally, in Section \ref{sec:experiments}, the impact of the use of provenance on space and time resources is measured experimentally. These results show that, provided a user accepts some performance trade-off, the library can provide articulate and intuitive results, when processors are composed to form complex computation chains.


\section{Related Work}\label{sec:related} 

Taken in a broad sense, we call ``data lineage'' any activity that attempts to link the result of a computation (its outputs) to elements that contribute to this result (its inputs). Depending on the field of study, variants on the notion of lineage have been given different names.

A large amount of work on lineage has been done in the field of databases, where this notion is often called \emph{provenance}. 
We can distinguish between three types of provenance. The first type is called \emph{why-provenance} and has been formalized by Cui \textit{et al.\@} \cite{DBLP:journals/tods/CuiWW00}. To each tuple $t$ in the output of a (relational) query, why-provenance associates a set of tuples present in the input of the query; the meaning of this set is to collect all the input data that helped to ``produce'' $t$. 
\emph{How-provenance}, as its name implies, keeps track not only of what input tuples contribute to the input, but also in which way these tuples have been combined to form the result \cite{DBLP:conf/pods/GreenKT07}. 
Finally, \emph{where-provenance} describes where a piece of data is copied from \cite{DBLP:conf/icdt/BunemanKT01}. It is typically expressed at a finer level of granularity, by allowing to link individual values inside an output tuple to individual values of one or more input tuple. 


There exist various implementations of provenance-aware database systems. Where-provenance has been implemented into Polygen \cite{DBLP:conf/vldb/WangM90}, DBNotes \cite{DBLP:conf/sigmod/ChiticariuTV05}, \textsc{Mondrian} \cite{DBLP:conf/icde/GeertsKM06}, MXQL \cite{DBLP:conf/icde/VelegrakisMM05} and \textsc{Orchestra} \cite{DBLP:conf/sigmod/KarvounarakisIT10}. The \textsc{Spider} system performs a slightly different task, by showing to a user the ``route'' from input to output that is being taken by data when a specific database query is executed \cite{DBLP:conf/vldb/ChiticariuT06}. The foundations for all these systems are relational databases, where sets of tuples are manipulated by operators from relational algebra, or extensions of SQL.

Outside the field of databases, the W3C has standardized a data model for provenance information called \textsc{Prov} \cite{w3c-prov}. The standard includes an ontology that defines  multiple provenance relationships, such as ``was derived from'', ``was revision of''. 
A templating system for \textsc{Prov} data has been proposed by Moreau \textit{et al.\@}\cite{moreau_templating_2018}; it resembles the graph of processors produced in the present work. However, \textsc{prov-template} assumes that, for a given processing task, this graph has the same structure for every input, and only differs in the actual bindings given to its various elements. On the contrary, we shall see that in BeepBeep, some processor chains produce graphs whose structure highly depends on the input given to the pipeline. Moreover, the approach assumes these templates as given, while our proposed work dynamically generates these graphs from a processor chain and an input stream at runtime.


On its side, dynamic taint analysis consists in marking and tracking certain data in a  program at run-time. TaintCheck is a system where each memory byte is associated with a 4-byte pointer to a taint data structure \cite{DBLP:conf/ndss/NewsomeS05}; program inputs are marked as tainted, and the system propagates taint markers to other memory locations during the execution of a program; this concept has been extended to the operating system as a whole in an implementation called Asbestos \cite{DBLP:journals/tocs/VandebogartEKKFZKMM07}. Hardware implementations of this principle have also been proposed \cite{DBLP:conf/asplos/SuhLZD04,DBLP:conf/micro/CrandallC04}. \textsc{Gift} is another taint analysis tool; Aussum is a compiler based on it \cite{lam_general_2006}.
Dytan \cite{DBLP:conf/issta/ClauseLO07}.
\textsc{Rifle} focuses on the information flow \cite{DBLP:conf/micro/VachharajaniBCROBRVA04}
TaintBochs is a system that has been used to track the lifetime of sensitive data inside the memory of a program \cite{DBLP:conf/uss/ChowPGCR04}.

On the stream processing front, few solutions have been developed to provide explanations for queries. Spline \cite{DBLP:conf/bigcomp/ScherbaumNV18} is a system that works on top of Apache Spark and attempts to recover lineage information by instrumenting processing jobs; ``lineage'', in this case, means the topological organization of jobs and data sources that are being used. However, this system does not work at the individual event level, and hence cannot be used to explain the value of a precise output event produced by a Spark pipeline. Apache Atlas\footnote{\url{https://atlas.apache.org}} provides similar coarse-grained functionaities for jobs running on Hadoop. To the best of our knowledge, no existing work focuses on fine-grained explainability of individual events in a stream processing pipeline.

\section{Event Log Query Processing with BeepBeep}\label{sec:bb} 

In this section, we shall first describe basic concepts of event log processing, as implemented by the BeepBeep event stream query engine. BeepBeep is a Java library that allows users to easily ingest and transform event streams of various types; the library is free and open source\footnote{\url{https://liflab.github.io/beepbeep-3}}. Over the past few years, BeepBeep has been involved in a variety of case studies \cite{DBLP:journals/cie/VarvaressosLGH17,DBLP:conf/fps/BoussahaKH17,DBLP:conf/isola/KhouryHW16,DBLP:conf/petra/HalleG016,iot-book}. 
A detailed description of BeepBeep is out of the scope of this paper, due to space restrictions. For further details, the reader is referred to a complete textbook describing the system \cite{beepbeep-book}.

\subsection{Functions and Processors}                      

BeepBeep is organized around the concept of \emph{processors}. In a nutshell, a processor is a basic unit of computation that receives one or more event streams as its input, and produces one or more event streams as its output. A processor produces its output in a streaming fashion: it
does not wait to read its entire input trace before starting to
produce output events. However, a processor can require more
than one input event to create an output event, and hence may
not always output something when given an input.
                                                                                            
\begin{figure}
\centering
\includegraphics[scale=0.42]{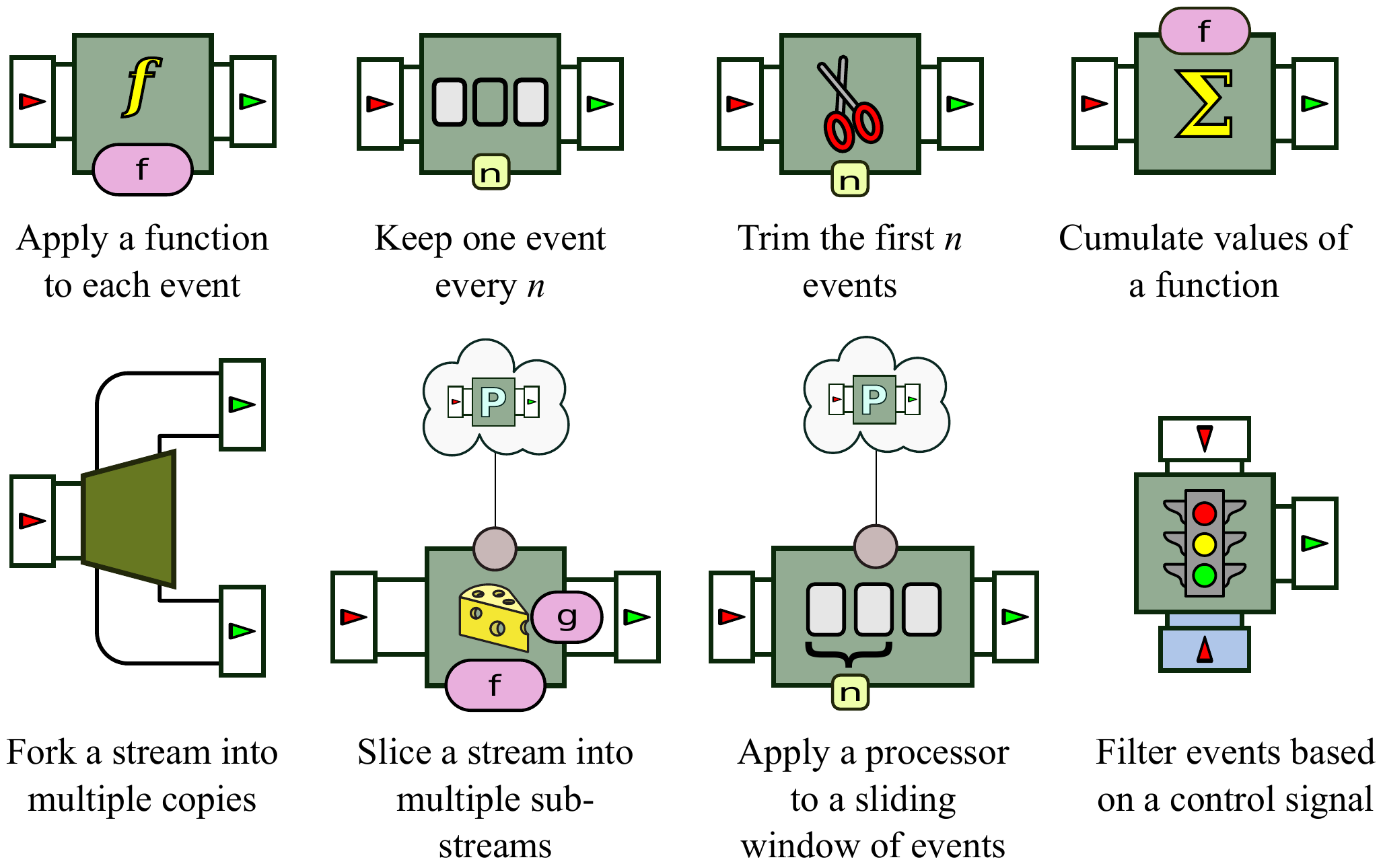}
\caption{Pictograms for the basic BeepBeep procssors.}
\label{fig:procs}
\end{figure}

BeepBeep's core library provides a handful of generic processor objects performing basic tasks over traces; they can be represented graphically as boxes with input/output ``pipes'', as is summarized in Figure \ref{fig:procs}.

A first way to create a processor is by lifting any function $f$ into processor. This is done by applying $f$ successively to each input event (or $n$-tuple of input events, for functions that have $n$ arguments), producing the output events. A variant of this process is the \texttt{Cumulate} processor, which, as its name implies, accumulates input values according to some function; for example, providing it with the \texttt{Addition} function will cause it to output the cumulative sum of all events received so far. Note that \texttt{Cumulate} also works with non-numerical events.

A few processors can be used to alter the sequence of events received. The \texttt{Count\-Decimate} processor returns every $n$-th input event and discards the others. Another operation that can be applied to a trace is trimming its output. Given a trace, the \texttt{Trim} processor returns the trace starting at its $n$-th input event. Events can also be discarded from a trace based on a condition. The \texttt{Filter} processor takes two input streams; the events are let through on its first input stream, if the event at the matching position of the second stream is the value \texttt{true} ($\top$); otherwise, no output is produced.

Another important functionality of event stream processing is the application of some computation over a window of events. If $\varphi$ is an arbitrary processor, the \texttt{Window} processor of $\varphi$ of width $n$ sends the first n events (i.e.\ events numbered 0 to $n-1$) to an instance of $\varphi$,
which is then queried for its $n$-th output event. The processor
also sends events 1 to $n$ to a second instance of $\varphi$, which is then also queried for its $n$-th output event, and so on. The resulting trace is indeed the evaluation of $\varphi$ on a sliding window of $n$ successive events. Any processor can be encased in a sliding window, provided it outputs at least $n$ events when given $n$ inputs.

In the case of business processes, a log can contain interleaved sequences of events for multiple process instances. The sub-sequence of events belonging to the same process instance is called a \emph{slice}; applying a separate processing to each such sub-sequence will be called \emph{slicing}. To this end, BeepBeep provides a processor called \texttt{Slice}, which is one of the most complex of the core library. It uses a function $f$ to separate an input stream into several sub-streams. Each of these sub-streams is sent to a different instance of some processor $P$, and the output of each copy is aggregated by another function~$g$.

\subsection{Pipes and Palettes}

In order to create complex computations, processors can be composed (or ``piped'') together, by letting the output of one processor be the input of another. An important characteristic of BeepBeep is that this piping is possible as long as the type of the first processor’s output matches the second processor’s input type. Such pipes can easily be created by using Java as the glue code.

If chains of basic processors are not sufficient to accomplish the desired computation, BeepBeep makes it possible to extend its core with various packages of domain-specific processors and functions, called \emph{palettes}. The main advantage of the palette system is its modularity: apart from a small core of common objects, a user is required to load only the palettes that are relevant to the computing task at hand. BeepBeep's ``standard library'' offers more than a dozen such palettes; we briefly describe in the following those of particular interest in the context of business process logs.

\subsubsection{Finite-State Machines}

A frequent use of stream processing is to check whether the events inside a log follow a specific sequence, and trigger a warning as soon as a violation is observed. Specifying the allowed event sequences can be done, among other things, by means of a finite-state automaton. BeepBeep's \palette{Fsm} palette allows users to create \emph{Moore machines}, a special case of automaton where each state is associated to an output symbol. This Moore machine allows its transitions to be guarded by arbitrary functions; hence it can operate on traces of events of any type.

By associating states of the FSM to, e.g.\ Boolean values, a Moore machine can act as a \emph{monitor}: when fed events from a log, it can be instructed to output the value \texttt{true} (or no value at all) as long as the input sequence is a valid path, and return \texttt{false} when the last event received does not correspond to an acceptable transition in the current state of the automaton.

\subsubsection{Linear Temporal Logic}

Similar to the \palette{Fsm} palette, the \palette{Ltl} palette makes it possible for users to write conditions on event sequences using Linear Temporal Logic (LTL) \cite{DBLP:conf/focs/Pnueli77}. We recall that LTL, in addition the usual Boolean connectives, provides four temporal \emph{operators} that apply on an arbitrary formula $\varphi$. The temporal operator {\bf G} means ``globally'': the formula $\mbox{\bf G}\,\varphi$ means that formula $\varphi$ is
true in every event of the trace. The operator {\bf F} means ``eventually''; the formula $\mbox{\bf F}\,\varphi$ is true if $\varphi$ holds for some future event of the trace. The operator {\bf X} means ``next''; it is true whenever $\varphi$ holds in the next event of the trace. Finally, the {\bf U} operator means ``until''; the formula $\varphi\,\mbox{\bf U}\,\psi$ is true if $\varphi$ holds for all events until some event satisfies~$\psi$.

Each of these temporal operators is implemented as a \texttt{Processor} object, and chaining such processors appropriately allows users to create pipes that can be used to evaluate any arbitrary LTL formula. 
Each LTL processor for an LTL formula $\varphi$ applies the following semantics: the $i$-th output event is the verdict produced by a monitor evaluating the input trace starting at event $i$.

Typically, temporal processors produce bursts of output events for multiple inputs at the same time, once a specific value (true or false) is received in the input stream. Consider the case of operator $\mbox{\bf G}\,\varphi$. The processor for this operator takes as input a stream of Boolean values, corresponding to the evaluation of $\varphi$ on each input event. Given the input stream $\top,\top,\bot,\top$, the processor will produce the output stream $\bot,\bot,\bot$: indeed, the property $\mbox{\bf G}\,\varphi$ is definitely false for the trace prefixes starting in each of the first three input events. However, those three outputs can only be produced once input event $\bot$ at position 3 has been received. Similarly, a definite verdict cannot yet be computed for the input prefix starting at event 4. A similar reasoning applies to the remaining operators.

\subsection{Examples}\label{subsec:bb-examples}

We now give a few examples of processor chains that can be built using the basic processors and the objects provided by the palettes just described. These examples are aimed at showing the diversity of computations that can be expressed with BeepBeep, and will be reused in the next section to illustrate how their results can be explained by the lineage tracking extensions introduced in this paper. They are by no means a complete showcase of BeepBeep's functionalities.

\subsubsection{Window Product}

\begin{figure}
\centering
\includegraphics[scale=0.3]{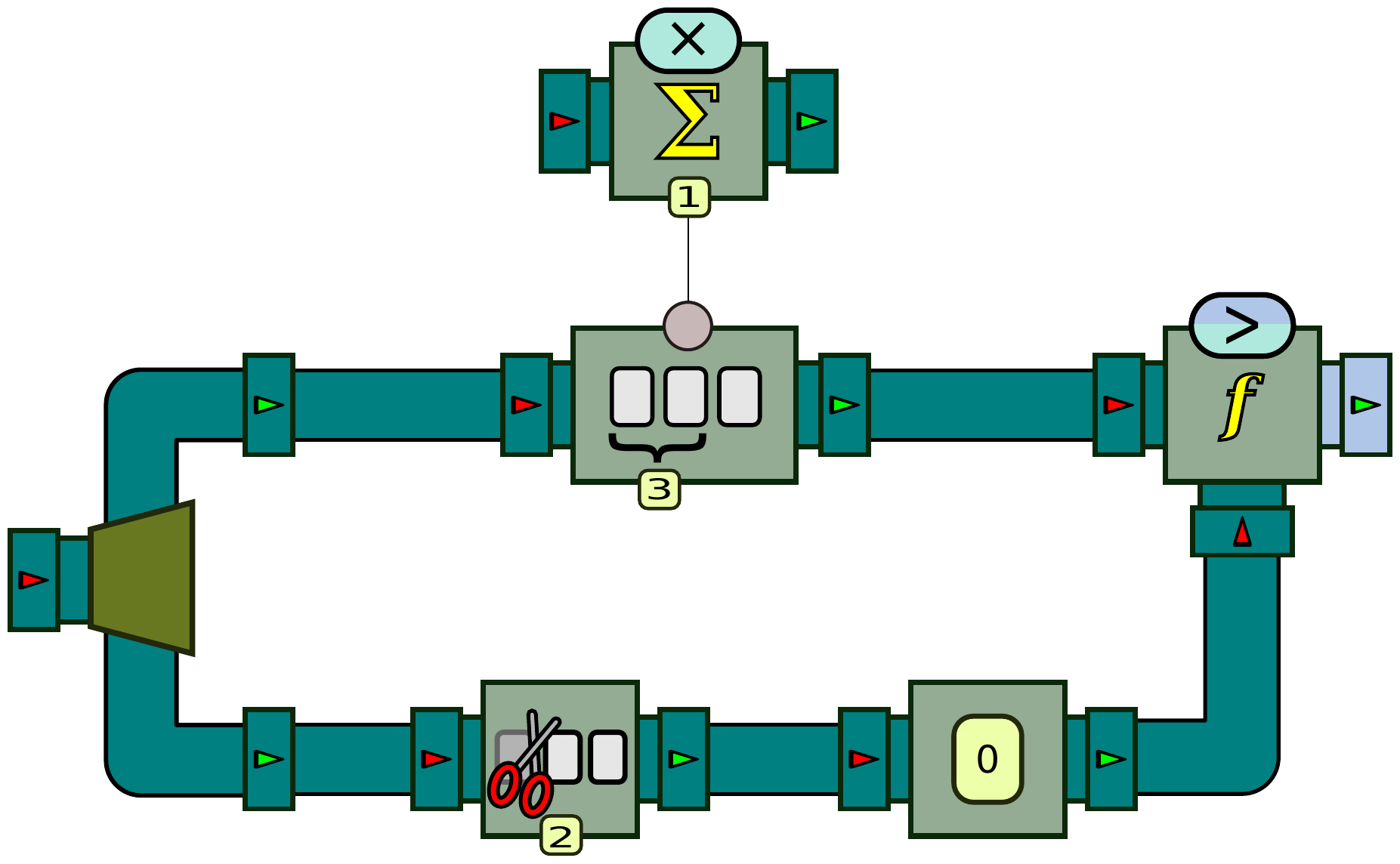}
\caption{The BeepBeep chain of processors for the \textit{Window product} query.}
\label{fig:product-gt-0}
\end{figure}

As a first example, consider the processor chain illustrated in Figure \ref{fig:product-gt-0}. This chain takes as input a stream of numerical values; it computes the product of each sequence of three successive values, checks whether this product is not equal to zero. This chain introduces a special processor, not described earlier, at the bottom of the figure, which simply turns any input event into a predefined constant ---in this case, the value 0.  Intuitively, the output of this chain can be translated as the assertion ``the product of any three successive values must be greater than zero''. Consider the input stream $3,1,4,0,5,9,2$ given to this pipeline. The output produced for this prefix will be the stream of Booleans $\top,\bot,\bot,\bot,\top$. Indeed, the first window of three events ($3,1,4$) has a non-null product; however, it is easy to see that the next three windows, which all contain the number $0$, have a product equal to zero and cause the emission of value~$\bot$.

\subsubsection{Process Lifecycle}

\begin{figure}
\centering
\includegraphics[scale=0.3]{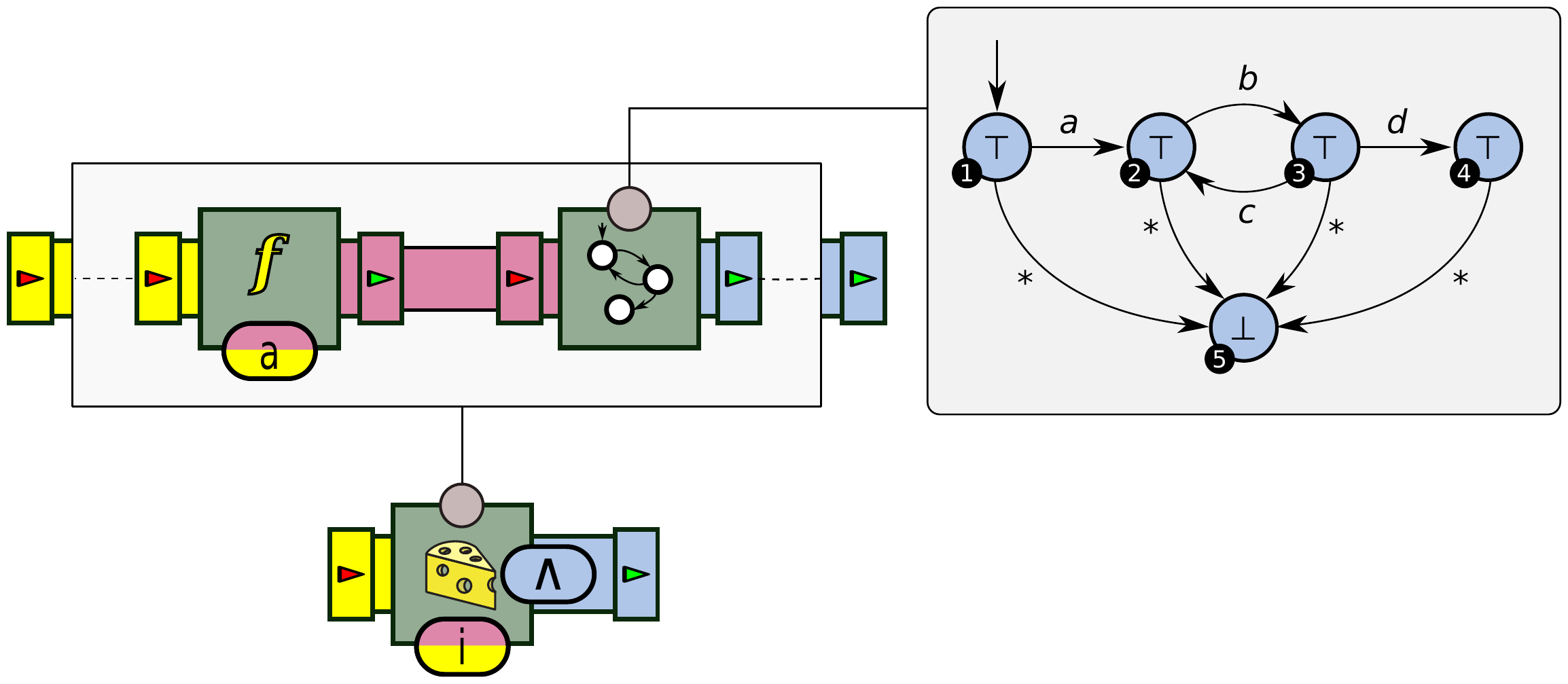}
\caption{The BeepBeep chain of processors for the \textit{Process lifecycle} property.} 
\label{fig:slice-fsm}
\end{figure}

A second example is shown in Figure \ref{fig:slice-fsm}. This time, input events are assumed to be tuples of the form $(i,a)$, where $i$ is some numerical identifier, and $a$ is the name of an action. This basic format is appropriate to represent a simple kind of business process log, where multiple interleaved process instances are distinguished by their value of $i$, and each instance is made of a sequence of actions. This use case is a prime example of the \texttt{Slice} processor, which in this case is used to separate events of each process instance based on their \texttt{id}, and feeds each sub-sequence into a chain that first fetches the \texttt{action} field of each event, and updates the state of a Moore machine accordingly.

In this particular case, one can see that the Moore machine for each instance has transitions to a ``sink'' state that produces value ``false'' ($\bot$). Any sequence that follows the intended pattern has the machine remain in a state that produces the value ``true'' ($\top$). Written as a regular expression, the language accepted by this machine corresponds to the string $a(bc)^+d$. The output of each Moore machine is aggregated into a Boolean conjunction; therefore, for the global processor chain to return $\top$, each currently active process instance must follow the intended lifecycle ---otherwise the chain returns $\bot$.

Consider for example the following sequence of actions: $(1,a),(2,a),(2,b),(1,b),(2,c),(2,d)$. The processor's output for this prefix will be the sequence of Booleans $\top,\top,\top,\top,\top,\bot$. As one can see, this sequence of events contains two interleaved process instances, labelled $1$ and $2$. The sequence of actions for process 1 follows the intended pattern ($ab$), while the sequence of actions for process 2 ($abcd$) violates the lifecycle on the last event.

\subsubsection{LTL Property}

Our last event log query involves Boolean connectives and LTL temporal operators. Its processor chain is shown in Figure \ref{fig:ltl-example}. In this case, we assume the input events are lines of a CSV file, each containing a tuple $(\mbox{\sl action},p)$, where \textsl{action} is an action name and $p$ is an arbitrary numerical value. The chain decomposes this tuple by fetching the value of $p$ (top branch) and the value of \textsl{a} (bottom branch). The condition $p<0$ is evaluated on the top branch; the condition $\mbox{\sl action}=a$ is evaluated on the bottom branch, for some predefined action name~$a$.

\newcommand{\ltlFormula}{\ensuremath{\mbox{\bf G}\, (p < 0 \rightarrow \mbox{\bf X}\,(\mbox{action} = a \wedge \mbox{\bf X}\, (\mbox{action} = a)))}}

\begin{figure*}
\centering
\includegraphics[scale=0.3]{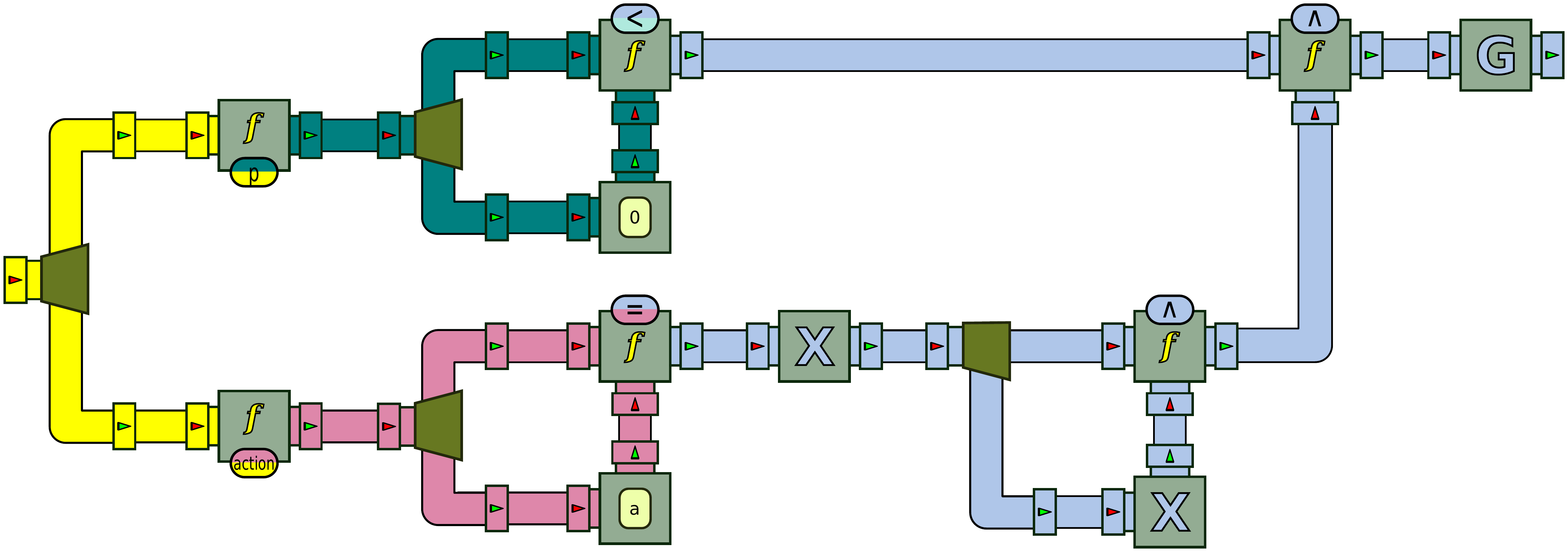}
\caption{The BeepBeep chain of processors that checks the LTL property \ltlFormula{}.}
\label{fig:ltl-example}
\end{figure*}

The Boolean streams corresponding to these conditions are then sent through a piping of Boolean connectives and LTL operators. The end result is also a Boolean stream, which amounts to the evaluation of the LTL formula \ltlFormula{}. Intuitively, this expression can be formulated as ``every input event with a negative value for $p$ must be followed by two successive events whose action is $a$''. The chain outputs $\bot$ whenever this pattern is not being followed in the input stream.

As an example, consider the input stream made of the following four tuples $(b,1),(c,-2),(a,0),(d,0)$. One can see that the output of the processor chain, after ingesting these four events, will be the sequence $\bot,\bot$. According to the semantics of LTL operators, this is caused by the fact that the sub-traces starting at the first and second event violate the condition expressed above: they both contain an event with $p<0$ that is not followed by two successive $a$. No definite verdict can be yet reached for the sub-traces that start at the third and fourth event; this is why no output event has been produced for these two inputs.


\section{An Explanation Mechanism for Stream Queries}\label{sec:lineage} 

After this brief presentation of the BeepBeep event stream library, we describe in this section how the original system has been retrofitted with data lineage functionalities. More precisely, in the present context ``lineage'' will correspond to the association that can be established between a specific output event produced by a processor, and the input events that are involved in the production of this output.

This is where BeepBeep's design principles, based on the concept of composition, can be put to good use. Since complex processor chains are obtained by piping basic processors into graphs, it suffices to define input/output associations for each processor separately. By virtue of composition, it will then be possible to retrace output events all the way up to the original inputs of a pipe, by simply following the chain of associations from each processor to the next upstream processor.

The goal of these additions and modifications is to make lineage as transparent as possible to the end user. The implications of this requirement are twofold. First, all modifications must preserve backward compatibility: existing programs using BeepBeep without lineage should still be valid programs under the new version. Second, benefiting from data lineage in a program should require as few modifications as possible to a processor chain; that is, lineage should come at a little cost in terms of added complexity to the glue code. The result of these modifications to the basic design of the library is described in what follows.

\subsection{The Event Tracker}

All data lineage functionalities in BeepBeep are centered around a singleton object called the \emph{event tracker}. The sole purpose of this object is to answer lineage queries: given an output event at a specific position in an output stream computed by a processor chain, the event tracker must point to the events of the chain's inputs that contribute to (or ``explain'') the fact that this particular output event contains this particular value.

In order to do so, the event tracker must be informed, by the various processors in the chain, of the output events they produce, and also to what input events they should be associated to. To this end, the \texttt{Event\-Tracker} interface declares a method called \texttt{associate()}, which can be called by processors during the execution of a task. A call to \texttt{associate()} must provide the following elements:
\begin{inparaenum}
\item The ID of the processor instance making the call
\item The index of the output pipe
\item The position of the output event in the output stream
\item The index of the input pipe
\item The position of the input event in the output stream
\end{inparaenum}.
As one can see, calls to this method can be used by implementations of \texttt{Event\-Tracker} in order to record input/output associations. Since each processor instance in BeepBeep is given a numerical identifier that is unique across a given program, the associations for each processor of a chain can be recorded and distinguished.

However, processors must be aware of the existence of such an event tracker so that they can call it. This is why the \texttt{Processor} class is modified in such a way that each of these objects can now store a reference to an event tracker. By default, lineage is turned off: processors are instantiated with a null reference as their default event tracker, indicating that no call to \texttt{associate()} needs to be made. This default can be changed by passing a non-null implementation of \texttt{Event\-Tracker} to a processor object after its creation.

Passing an event tracker to each processor instance one by one would be tedious; it would also violate our design principle of minimal modifications to the glue code. Since each processor in a chain is eventually connected to another one, an alternate approach is to use BeepBeep's \texttt{Connector} object, and arrange for the event tracker to be passed to processors through calls to \texttt{connect()}. In such a case, a user first instantiates a \texttt{Connector} by specifying an event tracker, and then uses this connector's \texttt{connect()} method to pipe processors, in place of the usual static method of the class. This call to \texttt{connect()} serves a double purpose: it makes processors aware of the existence of an event tracker, and it also allows the tracker to keep track of the connections between processors. Knowledge of these connections is necessary in order to follow lineage across the whole chain. Under such a design, a single line of glue code needs to be changed in order to enable lineage in a processor chain.

Once lineage has been properly set up in a program, a stream query can be evaluated in the usual way. At any moment during the processing, the event tracker can be asked for lineage information about a specific output event. This is done by calling a method named \texttt{get\-Provenance\-Tree()}. A provenance query contains three elements: the unique ID $n$ of a processor, the index $i$ of an output pipe, and the position $p$ of the output event in the corresponding output stream. Intuitively, such a query can be translated into the question: ``what is the explanation for the $p$-th event of the $i$-th output pipe of processor $n$?''

In return, the event tracker produces a directed acyclic graph (DAG) which, from the given output event, follows the input/output associations in the processor chain all the way up to the original inputs. As we shall see, the relationship between the input and the output can be many-to-many; this is why the generated structure is generally a graph, and not a linear chain of nodes.

\subsection{I/O Associations for Common Processors}

Equipped with this basic setup, supporting lineage in processors amounts to the insertion, in each class descending from the top-level \texttt{Processor}, of appropriate calls to a tracker's \texttt{associate()} methods. Since processors have a streaming mode of operation, these calls should also be made in a streaming fashion. This means that associations should be recorded progressively as the input events are ingested, as soon as such associations can be determined.


In general, all the inputs given to a computation are considered to explain the output; for example, with the function $f(x,y) = x+y$, one can see that any value $f$ can produce always depends on its two operands, $x$ and $y$. However, there exist exceptions to this general rule. Let us take the case of function $g(x,y) = xy$; typically, the knowledge of both $x$ and $y$ is required to explain the output value, but not always: when $x=0$ and $y=1$, the fact that $g(x,y)=0$ can be explained solely by the value of $x$. A similar argument could be done with Boolean connectives such as disjunction and conjunction.

In the following, we describe the rules used to produce input/output associations for the various functions and processor objects present in the BeepBeep library.

\subsubsection{Core Processors}

Most of BeepBeep's core processors have relatively straightforward association rules. The \texttt{Count\-Decimate} processor, whose task is to keep every $n$-th event and discard the others, registers an association between input event at position $i$ and output event at position $ni$. The \texttt{Trim} processor, which discards the first $n$ events, registers an association between input event at position $i$ and output event at position $i-n$ (for every $i \geq n$). The \texttt{Fork} processor simply replicates the input events to its outputs; the $i$-th input event is associated to the $i$-th output event of every output pipe.

The \texttt{Window} processor, which applies a processor $P$ on a sliding window of $n$ events, introduces a level of indirection. In order to produce the $i$-th output event from a stream of events $e_0, e_1, \dots$, the processor instantiates a copy of $P$ and feeds it with the interval of events $[e_i,e_{i+n-1}]$. It creates a temporary event tracker, instructed to intercept the input/output associations registered by $P$. From this tracker, the associations related to the last output event produced by $P$ are then transferred to the main event tracker, by taking care of shifting the positions of the input events by $i$. That is, the $k$-th event given to $P$ actually corresponds to the $(k+i)$-th event ingested by the \texttt{Window} processor.

These processors register the same associations, regardless of the actual content of the events they process. Some other processors will actually record different associations depending on the actual stream they receive. The I/O pairs for the \texttt{Apply\-Function} processor are determined by the I/O pairs of the underlying function that is being applied on each event front; as we have seen above, some of these functions may associate their output to all or part of their input arguments, depending on their values.

Similarly, the \texttt{Cumulate} processor generally associates the $i$-th output event to all input events up to the $i$-th: this is consistent with the fact that the processor computes the progressive ``accumulation'' of all input events received so far. However, this default behaviour may be overridden depending on the cumulative function being used. Take for example an instance of \texttt{Cumulate} processor applied on a stream of Boolean values, using logical conjunction as its function. On the input stream $\top,\top,\bot,\top$, the processor will return the output stream $\top,\top,\bot,\bot$ --that is, as soon as a false value is received, the processor's output will be false forever. To explain why a given output event at position $i$ is false, it suffices to point to an input event at position $j \leq i$ whose value is false.

Among all of BeepBeep's core processors, \texttt{Slice} is the one with the most complex I/O relationships. As a reminder, \texttt{Slice} creates multiple instances of a processor $P$, and dispatches an input event to an instance of $P$ based on the value returned by a slicing function $f$. The last output value produced by each instance of $P$ is then aggregated using another function $g$. Internally, each such copy of $P$ is linked to its own event tracker. To associate the $i$-th output event to inputs, the \texttt{Slice} processor first uses an internal event tracker to identify which of the arguments given to $g$ are involved in the production of its return value. These arguments correspond to output events produced by one or more instances of $P$; the event tracker for each of them is then queried in order to obtain the input events associated to that output event. Finally, as in the case of the \texttt{Window} processor, the relative event indices in each slice are converted into their corresponding positions in the stream ingested by \texttt{Slice}.

\subsection{I/O Associations for Palettes}

We shall now describe I/O associations that have been defined for processors of various palettes. As previously, our focus is on palettes that have particular relevance to the field of business processes.

\subsubsection{Moore Machines}

A Moore machine can be used to define compliance constraints related to the sequence of activities that can be seen in an instance of a process, in the form of a finite-state machine. When a violation to these compliance constraints is found in the log, existing tools, such as monitors, typically stop at the first event that makes the sequence non-compliant, and declare failure. The location in the trace where the monitor stops can already give some information to the user about the cause of the violation, but only in a fragmentary manner. Depending on the specification, the failure may be the result of the interplay between several events in the past that end up in a violation, and this information is not readily available by a classical monitor with a pass/fail verdict.

In order to address this issue, BeepBeep's \texttt{Moore\-Machine} processor has been retrofitted with lineage functionalities. Internally, each Moore machine instance records and updates a vector $\vec{v} = \langle (s_0, e_0, i_0), \dots, (s_n,e_n,i_n)\rangle$ whose elements are pairs $(s,e,i)$, where $s$ is a state of the machine, $e$ is an input event, and $i$ is the position of that event in the input stream. The vector is such that its last pair $(s_n,e_n,i_n)$ always contains the current state $s_n$ the machine is in. (If $\vec{v}$ is empty, the machine is in its initial state.)

Upon receiving an input event $e_{n+1}$, the machine updates this vector as follows. First, it takes the transition from its current state $s_n$, leading to a new state $s_{n+1}$. Assuming that $i_{n+1}$ is the number of input events received from the beginning of the stream, it then appends to the vector $\vec{v}$ the new triplet $(s_{n+1}, e_{n+1}, i_{n+1})$. The contents of this vector are then used to record associations between the $i_{n+1}$-th output event of the machine and its inputs; more precisely, the machine will register an association between the $i_{n+1}$-th output event and the $i_j$-th input event, for each $0 \leq j \leq n$. This corresponds intuitively to the fact that every input event in the vector is necessary in order to reach state $s_{n+1}$ and produce the corresponding output event. However, before moving on to the next input event, the Moore machine performs one last cleanup step. It looks for the earliest occurrence of $s_{n+1}$ in another triplet at some index $k \leq n$; if found, all the triplets at positions $j > k$ are deleted from the vector. 


The reason for this cleanup step is best explained on an example. Consider the Moore machine shown in Figure \ref{fig:slice-fsm}. Given the input sequence $\ev{a}_1\ev{b}_2\ev{c}_3$, the machine will produce the output sequence $\top_1 \top_2 \top_3$ (subscripts indicate event positions). According to the procedure just described, the third event of this output will be associated to the input events 1, 2 and 3. Suppose we now give the machine a new input event $\ev{b}_4$. In accordance to its transition relation, the machine will output a new symbol $\top_4$; however, the input associations for this symbol will be events at positions 1 and 4.

As the reader may have understood, the explanation produced for a given output event consists of a path from the initial state, excluding any loops that move away from a previously visited state. This corresponds to the intuition that the sequence of inputs $\ev{a}_1 \ev{b}_4$ suffices to produce $\top_4$. In other words, the machine finds the shortest subtrace in the input that produces the output.

This mechanism can be used to provide an explanation in the case of compliance violations. Suppose that in the previous example, state 5 corresponds to an error state. Therefore, an input sequence such as $\ev{a}_1 \ev{b}_2 \ev{c}_3 \ev{b}_4 \ev{c}_5 \ev{b}_6 \ev{a}_7$ violates the compliance requirement; however, in order to ``explain'' this violation, the subset $\ev{a}_1 \ev{b}_6 \ev{a}_7$ is sufficient.

\subsubsection{Linear Temporal Logic}

As we have seen, LTL is an alternate way in which compliance constraints on event sequences can be expressed. The \palette{Ltl} palette provides processors corresponding to each LTL operator, and equipped with lineage tracking functionalities. Their implementation is actually simpler than for Moore machines, and can be explained in a few words.

Consider the case of the processor for operator \textbf{G}. By virtue of the semantics of LTL, we know that this processor delays the production of output events as long as its inputs are true; once a false event is received, it produces a burst of $\bot$ output values. For a $\bot$ event that is emitted at position $j$, an association is recorded with the last input event at position $i \geq j$ whose value is false.

\begin{figure}
\centering
\includegraphics[width=2in]{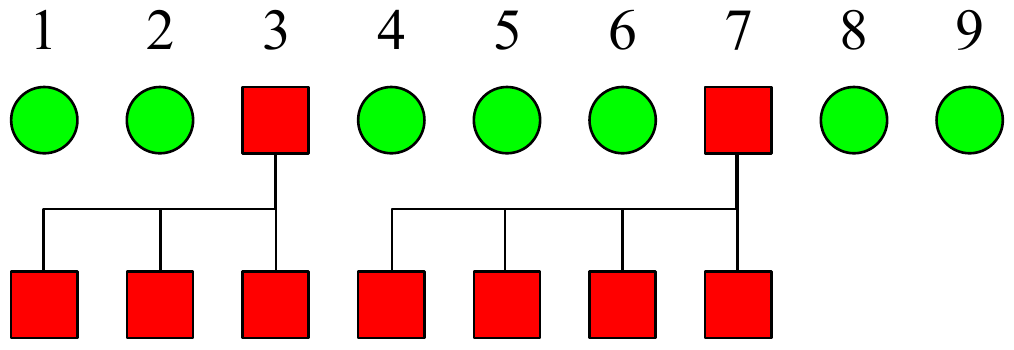}
\caption{Input/outputasssociations for the LTL \textbf{G} processor.}
\label{fig:ltl-g}
\end{figure}

This is illustrated in Figure \ref{fig:ltl-g}. The top row of the figure represents an input stream of Boolean values, with circles representing $\top$, and squares representing $\bot$. The bottom row shows the output produced by the \texttt{G} processor. Lines record the associations established between the inputs and the outputs. As one can see, the first three output events are associated with the first false value. Indeed, the verdict produced by the monitor for these three trace prefixes is ``caused'' by the presence of value $\bot$ at position 3. However, this event has no bearance on the output values produced for positions 4--7; they are rather caused by the presence of $\bot$ at input position 7. Hence, a temporal operator separates the output stream into zones, with each event of a zone typically being associated to the same event of the input stream. A similar reasoning can be applied to the other temporal operators.

\subsection{Examples}

These basic I/O associations turn out to provide surprisingly articulate and intuitive results, when processors are composed to form complex computation chains. 
We shall use the \textit{Window product} property to explain the operation of the event tracker. An explanation query is made of three elements:
\begin{inparaenum}
\item The ID of a processor in a chain;
\item The index of an output pipe on this processor;
\item The position of an event in the corresponding stream
\end{inparaenum}. From such a starting point, the event tracker will scan the input/associations recorded during the evaluation of a query, and recursively traverse these associations until the ultimate inputs of the chain are reached (or no upstream associations can be found to continue the chain).

\begin{figure}
\centering
\includegraphics[width=2in]{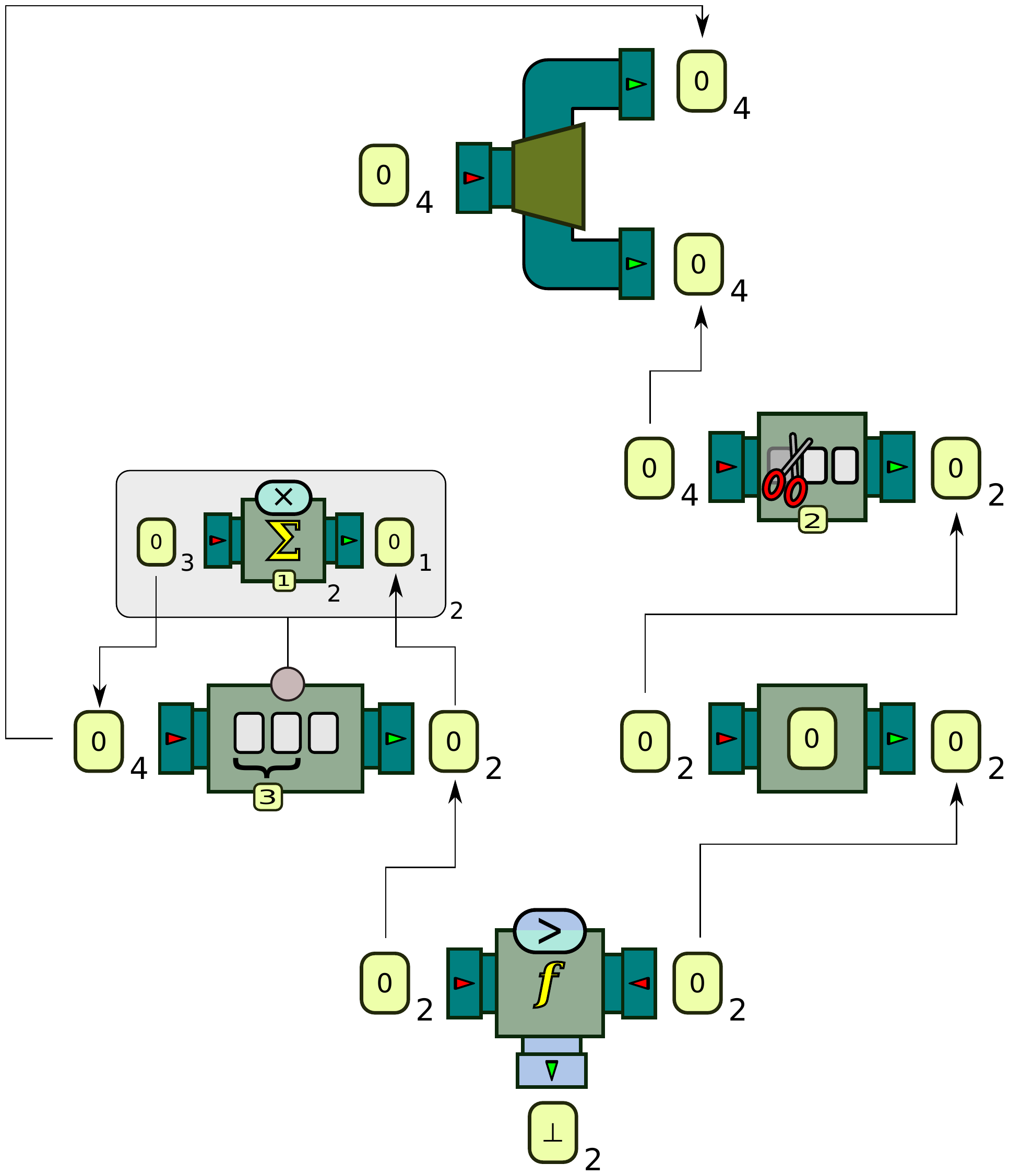}
\caption{An explanation graph for the \textit{Window product} query.}
\label{fig:explanation-graph}
\end{figure}

As we have seen earlier, on the input $3,1,4,0,5,9,2$, the \textit{Window product} processor chain produces the output $\top,\bot,\bot,\bot,\top$. Suppose we want an explanation for the reason the second event of this output is false. The \texttt{Event\-Tracker} associated to this processor chain is queried through a method called \texttt{get\-Provenance\-Tree()}, which will produce a directed acyclic graph whose structure is depicted in Figure \ref{fig:explanation-graph}. The graph is read from bottom to top; each input or output event is represented with a number corresponding to its relative position in the stream in question. Therefore, the direct explanation for the fact that processor $f$ returned $\bot$ on the second event is that it received $0$ as the second event in both its input streams. The chain can then be traversed further, and the reason for the production of each zero value can be retraced to different paths and input events in the processor chain.

Special attention should be given on the explanation for the result of the \texttt{Window} processor (left branch). This processor outputs a zero as its second event because the internal instance of the \texttt{Cumulate} processor associated to the second window returned zero. However, the reason for this null value is not explained by the whole window, but by the single 0 that corresponds, in this case, to the third event of the window. Ultimately, the whole graph converges back to a single input event, which is the zero value at position 4 in the input stream. This is in line with the intuition that output $\bot$ at position 2 is indeed caused by the presence of this zero in the input. Oftentimes, only the input/output associations of the extremities of the chain are relevant; in such a case, the graph can be ``flattened'' by keeping only the set of original input events that are mapped to a given output.

It is important to stress that this explanation graph depends on the output event chosen and the actual input stream given to the pipeline. Mere knowledge of the processor graph can be seen as lineage (similar to the information provided by Spline or Atlas), but is too coarse-grained to count as an \emph{explanation} of a result. Graphs of the same nature can be produced by the event trackers associated to the other processor chains illustrated in Section \ref{subsec:bb-examples}. They cannot be illustrated due to lack of space; however, the intuition behind them can be briefly discussed. In the case of the \textit{Process lifecycle} query, we have seen that the input stream $(1,a),(2,a),(2,b),(1,b),(2,c),(2,d)$ produces the $\bot$ output event at the sixth position, indicating that globally, not all process instances interleaved in the log are following the intended lifecycle. Again, the \texttt{Event\-Tracker} can be asked to explain this result. By following the I/O association rules for each processor in the chain, the end result will point to two events of the input log: tuples $(2,a)$ and $(2,d)$, corresponding to the second and sixth elements. This result provides two interesting pieces of information: first, the ID of the process that causes the global error, in this case process \#2. Second, the explanation mechanism identifies a minimal sub-trace for this process that causes the error. Here, we can see that in the complete trace $abcd$, the loop $bc$ has no impact on the erroneous result, and is therefore not included in the explanation.

Finally, a similar reasoning can be made on explanations for the third property, which involves LTL operators. It has been shown that the input sequence $(b,1),(c,-2),(a,0),(d,0)$ produces the output value $\bot$ at position 4. The explanation mechanism will retrace this output event to the inputs $(c,-2)$ and $(d,0)$. This corresponds to a ``witness'' of the fact that an event with $p<0$ has been seen, and that the second event that follows it does not have $a$ as its action. Notice how event $(a,0)$ is not part of the explanation, as it does not cause the erroneous verdict.


\section{Experimental Results}\label{sec:experiments} 

In order to assess the viability of such a system in practical situations, we performed an empirical evaluation of BeepBeep's lineage functionalities through an experimental benchmark. In this section, we report on these results, which have been obtained by running BeepBeep on various processor chains. They are aimed at measuring the impact, both in terms of computation time and memory, of the introduction of lineage functionalities inside the system. As we have seen, this is possible thanks to a switch provided by BeepBeep, and which allows users to completely disable lineage tracking if desired.


The experiments were implemented using the LabPal testing framework \cite{DBLP:journals/computer/HalleKA18}, which makes it possible to bundle all the necessary code, libraries and input data within a single self-contained executable file, such that anyone can download and independently reproduce the experiments. A downloadable lab instance containing all the experiments of this paper can be obtained from Zenodo, a research data sharing platform\footnote{%
The lab instance will be uploaded on Zenodo only for the final version of the paper. In the meantime, the latest version of the lab can be found on GitHub: \url{https://github.com/liflab/beepbeep-explainability-lab}%
}. 
All the experiments were run on a \machinestring{}, inside a Java 8 virtual machine with \jvmram{} MB of memory.



\subsection{Impact on Throughput}

The first element we measured is the impact on processing speed, or \emph{throughput}. Table \ref{tab:overhead-tp} shows the results for various types of stream queries. Each line represents a pair of experiments, corresponding to the evaluation of a stream query both with and without the use of a tracker. The measured value in each case is the average throughput, in number of input events processed per second.

\begin{table}
\centering
\scalebox{0.8}{\usebox{\ttCompTime}}
\caption{Relative throughput overhead.
}
\label{tab:overhead-tp}
\end{table}

Unsurprisingly, turning lineage on incurs a non-negligible slowdown, by as much as \maxOhThroughput{}$\times$ for the queries we considered. This is caused by the fact that, on each new event, a processor now calls the event tracker possibly multiple times, in order to register associations between inputs and outputs.

These results should be put in context with respect to existing works that include a form of lineage. The \textsc{Mondrian} system reports an average slowdown of 3$\times$ \cite{DBLP:conf/icde/GeertsKM06}; pSQL ranges between 10$\times$ and 1,000$\times$ \cite{DBLP:journals/vldb/BhagwatCTV05}; the remaining tools do not report CPU overhead. For taint analysis tools, Dytan reports a 30--50$\times$ slowdown \cite{DBLP:conf/issta/ClauseLO07}; GIFT-compiled programs are slowed down by up to 12$\times$; TaintCheck has a slowdown of around 20$\times$ \cite{DBLP:conf/ndss/NewsomeS05}, 1--2$\times$ for \textsc{Rifle} \cite{DBLP:conf/ndss/NewsomeS05}. Time overhead for Spline \cite{DBLP:conf/bigcomp/ScherbaumNV18} is close to zero, but as we have discussed, it provides lineage information at a much coarser level of granularity. Of course, these various systems compute different types of lineage information, but these figures give an outlook of the order of magnitude one should expect from such systems.

\subsection{Impact on Memory}

A second part of the experiment consisted in measuring the amount of additional memory required by the use of an event tracker. Memory was computed using the \texttt{Size\-Printer} object from the Azrael serialization library\footnote{\url{https://github.com/sylvainhalle/Azrael}}. This tool performs a recursive traversal of the member fields of a Java object, down to primitive types, and computes the sum of their reported sizes. The end result is a much more accurate indication of the memory actually consumed by an object, than would be a measurement of the JVM's memory footprint.

The results are summarized in Table \ref{tab:overhead-mem}. We can see that the relative impact on memory is larger than the impact of lineage on computation time. 
This is consistent with the intuition that lineage tracking requires one to ``remember'' more things, much more than to ``compute'' more things. This consumption is still relatively reasonable in the absolute: for example, with the \textit{Window product} processor chain, it would take an input file of \maxLines{} lines before filling up the available RAM in a 64 GB machine with lineage data.

\begin{figure}
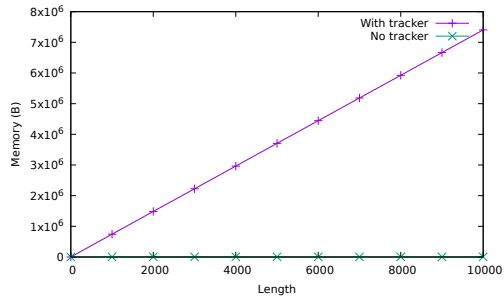

\centering
\scalebox{0.75}{\usebox{\pMemQueryWindowPproduct}}
\caption{Evolution of memory consumption during the evaluation of the \textit{Window product} query.}
\label{fig:comp-mem}
\end{figure}

The large relative blow-up is mostly caused by the fact that, for many processor chains, evaluating a query without lineage requires a constant amount of space, while the tracking-enabled pipeline uses a linear amount of space. This is illustrated in Figure \ref{fig:comp-mem}. As a matter of fact, it can be observed that for all the functions considered in this paper, each element of the output contributes for a constant amount of lineage data. Table \ref{tab:avg-oh} gives, for each query we considered, the average memory overhead per input event incurred by the use of an event tracker.
                                                          
\begin{table}
\centering
\scalebox{0.8}{\usebox{\ttCompMem}}
\caption{Relative memory overhead. 
}
\label{tab:overhead-mem}
\end{table}

\begin{table}
\centering
\scalebox{0.8}{\usebox{\tMemPerEvent}}
\caption{Average memory overhead (in bytes) per input event incurred by the use of an event tracker.}
\label{tab:avg-oh}
\end{table}

These figures should be put in context by comparing the overhead incurred by other lineage tracking tools. Notably, related systems for provenance in databases 
do not report their storage overhead for provenance data. Dynamic taint propagation systems report a memory overhead reaching 4$\times$ for TaintCheck \cite{DBLP:conf/ndss/NewsomeS05}, 240$\times$ for Dytan \cite{DBLP:conf/issta/ClauseLO07}, and ``an enormity'' of logging information for \textsc{Rifle} \cite{DBLP:conf/uss/ChowPGCR04} (authors' quote). Although these systems operate at a different level of abstraction, this shows that lineage tracking is inherently costly regardless of the approach chosen.


\section{Conclusion and Future Work}\label{sec:conclusion} 


In this paper, we have seen how an event stream processing engine called BeepBeep can be extended with functionalities for data lineage. In this particular context, lineage is the capability to link a part of the system's output all the way up to the concrete inputs that contributed to the production of that particular output. Thanks to BeepBeep's principle of composition, such lineage functionalities can be defined at the level of individual units of computation called \emph{processors}, whose input/output associations can then be chained to form a provenance graph. Through a few examples, it has been shown how such lineage capabilities can provide articulate and intuitive explanations for a result. What is more, those lineage functionalities are built-in, and transparent to the user: a single line of code suffices to switch the mechanism on or off. To the best of our knowledge, BeepBeep is the first event stream processing engine that provides such a simple, yet all-encompassing explanation system.

These promising results open the way to multiple research questions and improvements over this first solution. Extensions to BeepBeep have been developed to perform trend deviation detection and predictive analytics \cite{DBLP:conf/edoc/RoudjaneRKH19}, among other uses; it is planned to expand the basic explanation capabilities to these extensions in the near future. Currently, the system can only record associations between whole events. However, there exist situations where a finer granularity in the relationships between inputs and outputs would be required, such as when events are extracted from parts of a larger ``document'' such as an XML event. 

The implementation of the explanation mechanism could also be optimized in a few ways. First, we can observe that some processors always record the same association for each input/output event pair. 
Instead of recording this fact for every event, considerable savings, both in terms of time and space, could be achieved by making the tracker replace these individual associations with a single generic rule. Based on the promising results and the lessons learned from the implementation of BeepBeep's explanation mechanism, a redesign of the lineage functionalities is currently under way, and taking into account the previous observations.

The existence of a lineage tracking system inside BeepBeep  also opens the way to a myriad of exciting research questions. For example: 
For a given query, is there a part of the input event trace that never matters in the production of the output? Given that a part of the input is considered corrupted, are there parts of the output that are not affected by this corruption? What part of the input contributes the most to the output? All these questions could be studied both concretely (by studying a particular input-output pair), but more interestingly by reasoning over all the possible input-output pairs of a given processor chain.





\end{document}